\documentclass[aps,preprint]{revtex4}
\usepackage{amsmath}
\usepackage{graphicx}
\usepackage{amsfonts}
\usepackage{amssymb}
\begin{document}

\title{Demonstrating Multi-bit Magnetic Memory in the Fe$_{8}$ High Spin Molecule by Muon
Spin Rotation.}
\author{Oren Shafir and Amit Keren}
\affiliation{Physics Department, Technion-Israel Institute of
Technology, Haifa, Israel}
\author{Satoru Maegawa and Miki Ueda}
\affiliation {Graduate School of Human and Environmental Studies,
Kyoto University, Kyoto , Japan}
\author{Alex Amato, Chris Baines}
\affiliation{Laboratory for Muon-Spin Spectroscopy, Paul Scherrer
Institute, Villigen PSI, Switzerland}
\begin{abstract}
We developed a method to detect the quantum nature of high spin molecules
using muon spin rotation, and a three-step field cycle ending always with the
same field. We use this method to demonstrate that the Fe$_{8}$ molecule can
remember 6 (possibly 8) different histories (bits). A wide range of fields can
be used to write a particular bit, and the information is stored in discrete
states. Therefore, Fe$_{8}$ can be used as a model compound for Multi-bit
Magnetic Memory. Our experiment also paves the way for magnetic quantum
tunneling detection in films.
\end{abstract}

\maketitle

Immediately after the discovery of single molecular magnets it was suggested
that they could be used as multi-bit magnetic memory \cite{CaneschiJMMM99}.
However, this has never been demonstrated in the laboratory. For such a
demonstration the molecules must be subjected to several different magnetic
treatments (histories), which can then be clearly distinguished by a
measurement in a unique external condition. Other requirements from a good
magnetic memory is that, on the one hand a wide range of field will be
remembered in exactly the same way, and on the other hand, the stored
information will be discrete and hence very clearly distinguishable. Here we
demonstrate, using muon spin rotation ($\mu$SR) and a three-step field cycle
ending always with the same field, that the Fe$_{8}$ molecule can remember 6
(possibly 8) different histories and could serve as a model compound for
multi-bit magnetic memory.

Fe$_{8}$ is an abbreviation for [(C$_{6}$H$_{15}$N$_{3}$)$_{6}$Fe$_{8}$O$_{2}%
$(OH)$_{12}$]Br$_{7}$(H$_{2}$O)Br$\cdot$8H$_{2}$O, which was first synthesized
in 1984 \cite{WieghardtACIEE84}. The magnetically active part of this molecule
is constructed from 8 iron (III) ions, each having spin $S=5/2$. In the ground
state, 6 individual spins, out of the 8, point parallel to each other, while
the other 2 spins are directed anti-parallel to the first 6
\cite{PontillonJACS99,DelfsIC93}. As a result, Fe$_{8}$ molecules have a giant
electronic spin of $S=10$. In addition, the giant spins reside on a lattice,
and the interaction between them are small compared to the intramolecular
interactions \cite{FurukawaPRB02}. Due to the anisotropy of the crystal field,
the main terms of their Hamiltonian in the presence of an external magnetic
field $H$ in the $\widehat{\mathbf{z}}$ direction is $\mathcal{H}=-DS_{z}%
^{2}-g\mu_{B}S_{z}H$, where $D=0.27~$K \cite{SangregorioPRL97,BarraEL96},
$g=2$ is the spectroscopic splitting factor, and $\mu_{B}$ is Bohr magneton.
The most outstanding feature of this Hamiltonian is that, at certain
\textquotedblleft matching\textquotedblright\ fields, given by $H_{m}%
(n)=nD/g\mu_{B}=n\times2.2$~Oe where $-10\leq n\leq10$ is an integer, states
with positive and negative $S_{z}$ values have identical energies
\cite{DelfsIC93,BarraEL96}. As a result, at the matching fields tunnelling can
take place between spin states of opposite polarization. Similarly, when $H$
is very different from $H_{m}(n)$ these transitions are suppressed. This
property can be utilized to prepare the system in a number of different spin
configurations, which are stable for a relatively long time, by cycling an
external magnetic field, even though at the end of the cycle the field returns
to the same value, which is nearly zero field.

For the preparation of various spin states we start by applying a strong
negative (or positive) magnetic field of -10~kOe (10~kOe) parallel to the easy
axis for 10 minutes to polarize the Fe$_{8}$ molecules. In such a strong field
all the spins are in their $S_{z}=-10$ ($10$) state. The field is then swept
to an intermediate positive (negative) value $H_{i}$, at a rate of 41~kOe/sec,
crossing a matching field in the process, and, as a consequence, populating
states with positive (negative) $S_{z}$. Finally, the field is swept back to
+50~Oe (again crossing several matching fields), where the changes in the
sample magnetization as a function of $H_{i}$ are determined.%

\begin{figure}
[ptb]
\begin{center}
\includegraphics
{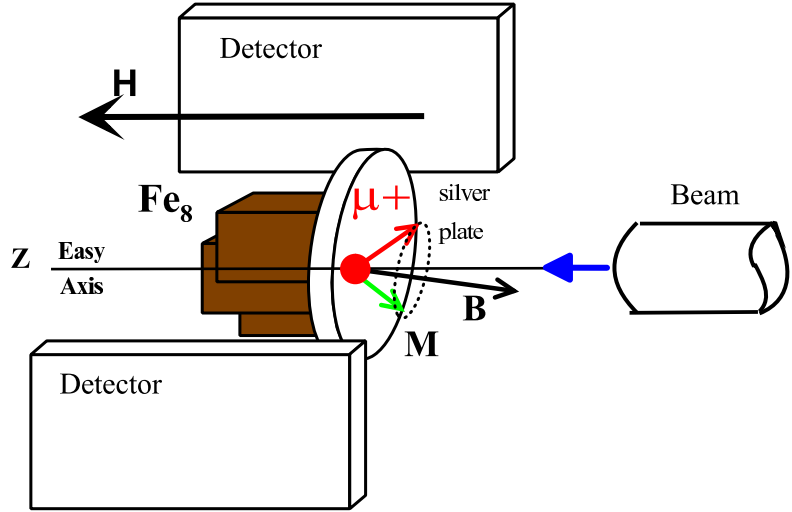}%
\caption{An \emph{indirect} $\mu$SR experimental setup showing the mosaic of
single crystals, and the external field ($H$) and the beam directions, which
are both in the molecules $\widehat{\mathbf{z}}$ direction. $\mathbf{M}$ is a
possible field at the muon site due to the molecules, and $\mathbf{B}$ is the
total field around which the muon spin rotates. The positron detectors are
also presented.}%
\label{MuSR}%
\end{center}
\end{figure}

The determination of the magnetization was done using the muon
spin rotation [$\mu$SR] technique \cite{BrewerMS98} and was
performed at the low temperature facility (LTF) of the Paul
Scherrer Institute, Switzerland. This technique is used to study
the magnetic properties of materials through direct measurement of
the time dependence of a positive muon ($\mu^{+}$) polarization,
and it is sensitive to field changes on the order of $1~$G. The
essence of the $\mu$SR measurement is that the muon spin rotates
around the axis of the internal field, and decay positrons are
emitted asymmetrically with respect to the muon spin direction.
Positron detectors are placed perpendicular to $\widehat
{\mathbf{z}}$, the direction of the external field and beam, as
depicted in fig.~\ref{MuSR}. In this figure one of the possible
directions of the magnetization and the internal field $B=H+4\pi
M$, around which the muon spin rotates, is shown. The initial
polarization of the muons was set to be 45$^o$ 
 relative to
$\widehat{\mathbf{z}}$. The data from the two counters are used to
reconstruct the asymmetry of the muon decay as a function of time
from the moment the muon arrives. In our experimental
configuration, this asymmetry is proportional to the muon spin
polarization in the direction perpendicular to
$\widehat{\mathbf{z}}$. Monitoring the muon polarization rotation
frequency for various $H_{i}$ allows us to detect clearly the
different spin configurations. Preliminary data using this method
was presented in Ref.~\cite{SalmanPhysicaB03}.

Our $\mu$SR measurements are performed on a few Fe$_{8}$ single crystals,
glued on a small silver plate (0.5mm thick) sample holder using GE varnish,
and aligned so that their direction (easy axis \cite{UedaJPSJ01}) is parallel
to the muon beam direction and external field directions. The Fe$_{8}$ single
crystals were positioned in two different configurations. In the first
configuration (indirect) the plate faced the beam, meaning the muons hit the
plate and are at best 0.5~mm away from the crystals. This configuration is
shown in Fig.~\ref{MuSR}. In the second configuration (direct) the Fe$_{8}$
faced the beam, meaning most of the muons hit the Fe$_{8}$ but some miss the
crystals and land on the silver plate or GE varnish very close to the
crystals. We tested both configurations since it was not clear, \emph{a
priori}, which one would yield better results. For each configuration a
calibration test was performed (no-Fe$_{8}$) in which the Fe$_{8}$ was removed
from the holder and the muons stopped in the silver plate. This test was
necessary because a superconducting magnet, like the one used in the
experiment, could have trapped flux, leading to a slightly different measured
field for different intermediate fields, even though the magnet power supply
was set to produce $+50~$Oe at the end of each field cycle. In both
configurations the samples were cooled down to a temperature of 40-100 mK in
order to minimize the thermal activation effects (the tunnelling process
dominates below 360mK \cite{SangregorioPRL97,OhmLTP98}).%

\begin{figure}
[ptb]
\begin{center}
\includegraphics
{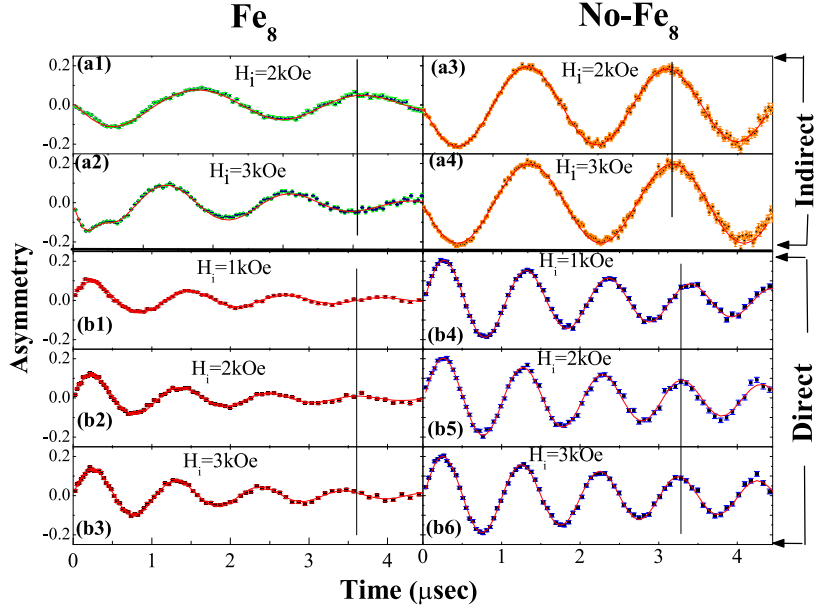}%
\caption{The muon asymmetry as a function of time from a sample of Fe$_{8}$ or
empty silver sample holder and different intermediate fields $H_{i}$, for both
indirect configuration where muons hit the silver sample holder only (a1-a4)
and direct configuration where muons hit mostly the Fe$_{8}$ (b1-b6). The
different intermediate fields ($H_{i}$) lead to different muon rotation
frequencies when Fe$_{8}$ is included (a1-a2 and b1-b3), but to similar
frequencies in an empty holder (a3-a4 and b4-b6).}%
\label{RawData}%
\end{center}
\end{figure}

In Fig.~\ref{RawData} we present the asymmetry as a function of time for both
direct and indirect configurations, for different intermediate fields $H_{i}$,
and the calibration. The results of the indirect experiments are depicted in
the four top panels (a1-a4) while the results of the direct experiment are
shown on the bottom six panels (b1-b6). All panels on the left side are
measurements with the Fe$_{8}$ and on the right are calibration measurements
of the empty silver sample holder. The most noticeable variation in
Fig.~\ref{RawData} is the difference in amplitude between the experiment with
the Fe$_{8}$ and the calibration experiment with an empty holder. This is due
to quick depolarization outside the $\mu$SR time window of some of the muons
stopping at sites with strong magnetic fields. Due to this depolarization we
were sceptic about the success of the direct experiment. This fast
depolarization is discussed further below. A more important but subtle change
in the data is the difference in the muon asymmetry between runs with the
Fe$_{8}$, panels (a1) to (a2) and (b1) to (b3). The different intermediate
fields cause different precession frequencies of the muon spin. This is
emphasized in the direct configuration by the vertical solid line passing
through the fourth maximum of the $H_{i}=2~$kOe run [panel b2]. For
comparison, the fourth maximum of the $H_{i}=1~$kOe and 3~kOe [panels (b1) and
(b3)] are a quarter wave to the left and to the right of this line,
respectively. In contrast, the situation in the calibration case, panels (b4)
to (b6), is different. There is no noticeable difference in the muon asymmetry
between the intermediate fields $H_{i}=2~$kOe and 3~kOe [panels (b5) and
(b6)], and only a small difference between the $H_{i}=1~$kOe and $2$~kOe
measurements [panels (b4) and (b5))]. Therefore most of the frequency shift is
due to the Fe$_{8}$ molecules. A similar demonstration is presented by the
vertical solid line in the indirect experiment. In this measurement two
frequencies showed up in the muon precession when the Fe$_{8}$ is in use. The
reason for having two frequencies is not yet clear to us. It should be pointed
out that the difference in muon precession frequency between the direct and
indirect experiments in the No-Fe$_{8}$ case is due to different plate
position with respect to the center of the magnet.%

\begin{figure}
[ptb]
\begin{center}
\includegraphics
{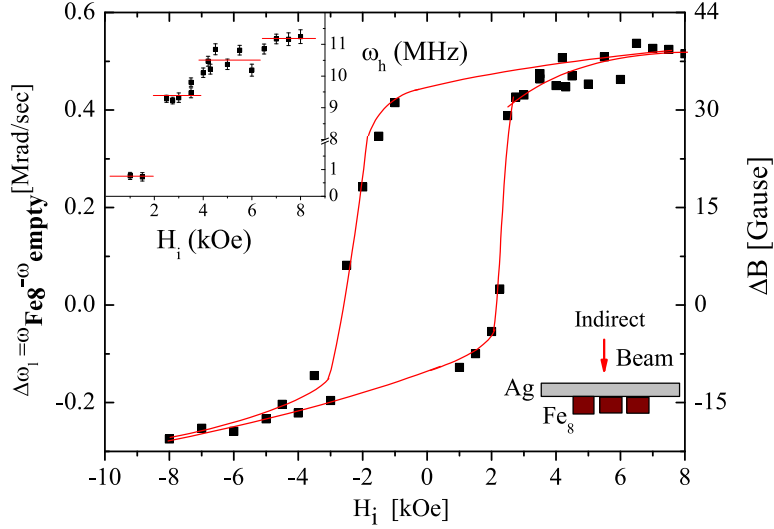}%
\caption{The shift in the muon low rotation frequency $\omega_{l}$ as a
function of the intermediate field $H_{i}$, between an empty sample holder,
and a holder with Fe$_{8}$ in the case of indirect measurement. The high
frequency component $\omega_{h}$ versus $H_{i}$ is shown in the inset. The
solid lines are guides to the eye.}%
\label{Hysteresis}%
\end{center}
\end{figure}

The data are analyzed by fitting the asymmetry of all measurements with a
function of the form:%
\[
A(t)=A_{h}\exp(-\lambda_{h}t)\cos(\omega_{h}t)+A_{l}\exp(-\lambda_{l}%
t)\cos(\omega_{l}t)
\]
where $\omega=\gamma_{\mu}B$ is the muon frequency, $B$ is the magnetic field
experienced by the muon, $\gamma_{\mu}=2\pi\cdot135.54$~MHz/T is the muon
gyromagnetic ratio, and $\lambda$ is the decay coefficient. The subscript $h$
($l$) stands for high (low) frequency and high (low) relaxation. In the direct
experiment a single component oscillating function was sufficient for the fit.

In Fig.~\ref{Hysteresis} we present the indirect experimental results. In this
figure, the low frequency shift $\Delta\omega_{l}$ between the calibration
test and the experiment with the sample is plotted as a function of the
intermediate field $H_{i}$ and the solid line is a guide to the eye. This
solid line forms an unusual kind of hysteresis loop despite the fact that the
measurements were always done in the same conditions. The shift saturates at
high intermediate fields, and it changes drastically in the regime of
$H_{i}\approx\pm2$~kOe. In addition, there is a big difference between
$\Delta\omega_{l}$ of intermediate fields with opposite sign, namely, the
hysteresis loop is not symmetric around zero shift. Some of the high
frequencies $\omega_{h}$ as a function of the intermediate fields are shown in
the inset of Fig.~\ref{Hysteresis}. Despite the noisy data, there is some
indication of steps, which are emphasized by the horizontal lines.

\begin{figure}
[ptb]
\begin{center}
\includegraphics
{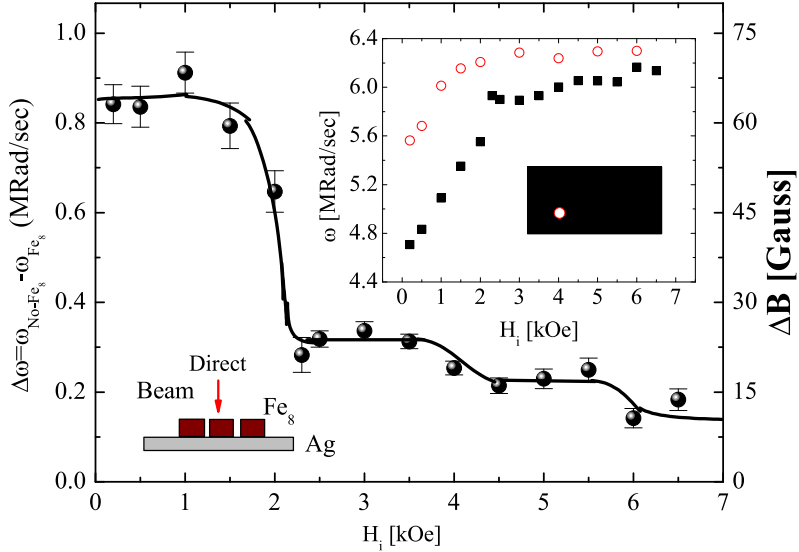}%
\caption{The shift of the muon rotation frequency as a function of the
intermediate field $H_{i}$, between the measurement of the empty silver sample
holder and that of a holder with Fe$_{8}$, in a direct measurement. The solid
line is a guide to the eye. Raw data of $\omega$ versus $H_{i}$ with, and
without, Fe$_{8}$ are shown in the insert.}%
\label{Steps}%
\end{center}
\end{figure}

A better frequency resolution was achieved in the direct experiment presented
in Fig.~\ref{Steps} where the frequency shift $\Delta\omega$ between the
calibration test and the sample measurements is presented. Again, the solid
line is a guide to the eye. Since only one oscillating component similar to
$\omega_{l}$ was required for the fit in this case, $\Delta\omega$ has no
subscript. The shift is the largest at low $H_{i}$ and decreases sharply when
$H_{i}$ grows toward the first matching field of $\sim$2.2~kOe. Upon further
increase of $H_{i}$, $\Delta\omega$ stays flat until a second decrease occurs
slightly above 4~kOe. This situation seems to repeat itself above 6~kOe as
well. We did not apply negative $H_{i}$ in the direct experiment.

It is important to note that these measurements have been repeated more than
once in order to check whether the results are reproducible. Performing the
same field cycle twice always gave exactly the same muon behavior in both
sample and calibration measurements. This provides reassurance that the
frequency shift is reproducible. Also, in other magnetization experiments
\cite{WernsdorferS99} the biggest change in the magnetic moment is found at
the second matching field $H_{m}(n=2)$ and not in the first one $H_{m}(n=1)$.
The fact that we detect the biggest shift at $H_{m}(n=1)$ in both of the
experimental configurations (Fig.~\ref{Hysteresis} and Fig.~\ref{Steps}) could
be explained by non-perfect alignment of the Fe$_{8}$ mosaic resulting in a
field perpendicular to $\widehat{\mathbf{z}}$. An alternative explanation is
avalanche of the magnetization, which are known to occur in large crystals. In
both cases most of the magnetization flips at the $H_{m}(n=1)$ leaving little
magnetization to vary at a higher matching field, and reduced sensitivity of
the shift to higher $H_{i}$. However, in principle, once films become
available, avalanches might not be a problem and one might be able to tune the
height of the jumps by a transverse field, leading to the different steps
having equal sensitivity.%

The dependence of $\Delta\omega$ on $H_{i}$ is the main finding in this work.
First of all, it reveals once again the quantum nature of these crystals,
since the period of $\Delta\omega(Hi)$ seems to agree with the $2.2$~kOe
period found by other methods \cite{WernsdorferS99}. Secondly, and more
importantly, it shows that Fe$_{8}$ molecules can \textquotedblleft
remember\textquotedblright\ which intermediate field was visited with 3 clear
plateaus (perhaps 4) and also can distinguish between positive and negative
$H_{i}$, giving all together 6 (possibly 8) different memory bits. In
addition, a field range of 2~kOe can be used to write a particular bit, but
the stored information is discrete. Finally, this memory lasts at least on the
time scale of the $\mu$SR measurement (1/2 hour). In fact, we have preliminary
data showing that this memory lasts at least for several hours (not shown). It
is this recollection of $H_{i}$, and the fact that many final frequencies can
be prepared, which warrants Fe$_{8}$ molecules the candidacy for a multi-bit
magnetic memory model compound.

Next, we address the question of the decrease in asymmetry between the sample
and holder measurements. We noticed, by measuring samples of increasing size,
that the amplitude of the muon rotation decreased. This lead us to a
speculation that muons that enter the sample relax immediately, and we are
actually observing only the muon that missed the sample and landed in the
sample holder or GE varnish. Since Fe$_{8}$ is a ferromagnet, the field it
creates outside the crystal depends on the average polarization of the
molecules, and it is this average that we detect.%

\begin{figure}
[ptb]
\begin{center}
\includegraphics
{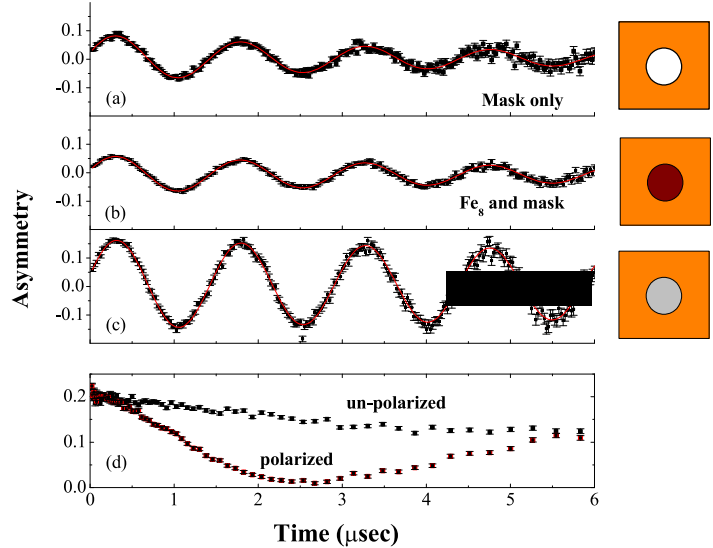}%
\caption{A demonstration that Fe$_{8}$ depolarize muons immediately. The
asymmetry of a hematite and glue mask (a) is very similar to mask and Fe$_{8}$
(b), but different from mask and silver (c). Therefore, muons in Fe$_{8}$ do
not contribute asymmetry (see text). In (d) we demonstrate that unpolarized
Fe$_{8}$ sample has very small muon relaxation. Therefore the signal is from
muons that experience the averaged field which is zero only outside the
sample.}%
\label{test}%
\end{center}
\end{figure}

In order to check this hypothesis we measured the asymmetry of three different
samples: (I) a mask made of hematite and glue with a hole in its centre, using
a veto counter (muons that missed the mask are not counted). In this case we
found asymmetry $A_{0}=0.084$ (see Fig.~\ref{test}a); (II) the same sample as
in (I) but having the hole covered with Fe$_{8}$ powder (see Fig.~\ref{test}%
b). In this case we found $A_{0}=0.067$; (III) the same as in (I) but having
the hole covered with silver (see Fig.~\ref{test}c). In this case
$A_{0}=0.157$. From (I) and (II) we can write%
\[
A_{0}^{Fe_{8}}P_{h}+0.084P_{m}=0.067
\]
where $A_{0}^{Fe_{8}}$ is the asymmetry from Fe$_{8}$, and $P_{m}$ and $P_{h}$
are the probabilities of landing in the mask or hole respectively. From (III)
we get%
\[
0.24P_{h}+0.084P_{m}=0.157
\]
since in silver $A_{0}=0.24$. The solution of these equations, together with
$P_{h}+P_{m}=1$, gives $A_{0}^{Fe_{8}}=0.047$. When considering the fact that
in our shift experiment the Fe$_{8}$ crystals are also covered with a thin
layer of GE varnish, $A_{0}^{Fe_{8}}$ is consistent with zero. We therefore
conclude that muons in Fe$_{8}$ relax immediately as they do in hematite.

Another indication to support this conclusion is shown in Fig.~\ref{test}d.
Here we present the muon longitudinal polarization (in the beam directions) in
zero field before and after the Fe$_{8}$ electronic spins have been polarized.
In the unpolarized state the muon relaxation is weak. Therefore, all muons
experience a weak field. This signal cannot come from muons inside the sample
where they interact with their neighboring spins and the average of the field
magnitude is not zero. This indicates that the signal is from muons outside
the sample. Indeed, when the Fe$_{8}$ spins are polarized the muon relaxation
increases, since the field strength outside the sample increases.

This experiment indicates that in our shift measurement we detect muons that
are stopped either in the thin layer of GE varnish covering the crystals or on
the silver plate in between crystals. These muons will experience an averaged
field from many molecules.

Finally, molecular clusters have been widely investigated as a model for
magnetism at the nano-scale, especially for quantum tunnelling of the
magnetization (QTM), since the memory stored in a single molecule can be lost
via QTM without thermal assistance. In parallel, a considerable effort is
being made to produce films of molecular magnets
\cite{CaroCM00,CaroCVD02,CasellasSP03}. Once this goal is achieved it will be
essential to probe their magnetization. This is not a trivial matter since
many experimental techniques that probe bulk materials are not applicable to
films. Our work presents one way to probe QTM in this situation since the
recent development of the slow muon apparatuses at PSI allow us to stop muon
in films of various thicknesses \cite{BakuleCP04}. We learned here that it
would be best to grow the films on a non magnetic substrate and to implant the
muons in the substrate rather than in the film. The quantum nature of the
films could then be investigated using the three step field cycle.

This work was performed at the Swiss Muon Source, Paul Scherrer Institute,
Villigen, Switzerland, and supported by the Swiss Federal Office for Education
and Science (BBW). We are grateful to the machine and instrument groups whose
outstanding efforts have made these experiments possible. It was also
supported by Israeli Science foundation under the Bikora program.

\end{document}